\documentclass[showpacs,preprintnumbers,aps,nofootinbib,floatfix,twocolumn]{revtex4}

\usepackage{epsfig}
\usepackage{citesort}

\def  \meee   {$\mu^- \to e^+ e^- e^-$ }
\def  \beq    {\begin{equation}}
\def  \eeq    {\end{equation}}

\def  \beqa   {\begin{eqnarray}}
\def  \eeqa   {\end{eqnarray}}

\def  \bcen   {\begin{center}}
\def  \ecen   {\end{center}}

\begin{document}

%%%%%%%%%%%%%%%%%%%%%%%%%%%
%  Titlepage

%\begin{titlepage}
%\renewcommand{\thefootnote}{\fnsymbol{footnote}}

%\vspace*{-0.5truecm}
%\begin{flushright}
%{\tt hep-ph/0508146 }
%\end{flushright}
%\vspace*{0.5truecm}

\title{Lepton number violation in Little Higgs model}

\author{S. Rai Choudhury} 
  \email{src@physics.du.ac.in}
\author{Naveen Gaur}
   \email{naveen@physics.du.ac.in}
\author{Ashok Goyal}
   \email{agoyal@iucaa.ernet.in}
\affiliation{Department of Physics $\&$ Astrophysics,
University of Delhi, Delhi - 110 007, India.}

\date{\today}
\begin{abstract}
In this note we examine the constraints imposed by muon
anomalous magnetic moment ($(g-2)_\mu$) and \meee on lepton number
violating (LNV) couplings of the triplet Higgs in Little Higgs (LH)
model. 
\end{abstract}

\maketitle

%%============================================================%%
%% Section 
%\section{Introduction \label{section:1}}

The Standard Model (SM) has been remarkably successful in explaining
experimental data upto the highest energies available at present. The
precision electroweak data suggests that the Higgs boson remains
light \cite{Eidelman:2004wy}, $m_H < 219$ GeV at 95\% CL, upto the
Planck's scale. In SM, 
the Higgs boson however, gets 
quadratically divergent contribution to its mass and requires fine
tuning of parameters which are sensitive to new physics that may be
present at scales much higher than electroweak scale. Fine tuning and
naturalness requires this new physics to be at the TeV
scale. Supersymmetry (SUSY) provides a particularly elegant solution
to the Hierarchy problem where quadratic divergences in Higgs mass are
canceled between contributions of SM particles and their
superpartners. This has the desired effect of stabilizing the
electro-weak scale. In Technicolor theories, the hierarchy problem is
deferred by introducing new dynamics at a scale not too much above
electroweak scale. Theories of large extra dimensions resolve the
hierarchy problem by lowering the Planck's scale and 
modifying quantum gravity at the TeV scale. Phenomenological
consequences of these theories have been studied in the literature
and constraints obtained \cite{Arkani-Hamed:1998nn}. 

\par Recently there has been a proposal to consider Higgs fields as
pseudo-Nambu-Goldstone boson of a Global symmetry which is
spontaneously broken at some high scale
\cite{Arkani-Hamed:2001nc}. The Higgs fields acquire mass 
through electroweak symmetry breaking triggered by radiative
corrections leading to Coleman-Weinberg type of potential. Since the
Higgs is protected by approximate global symmetry, it remains light
and the quadratic divergent contributions to its mass are canceled by
the contributions of heavy gauge bosons and a heavy Fermionic state
that are introduced in the model. The Littlest Higgs (LH) 
\cite{Han:2003wu,Han:2005dz,Arkani-Hamed:2001nc}
model is a
minimal model of this class which accomplishes this task to one loop
order within a minimal matter content. The LH model
consists of an 
SU(5) non-linear sigma model which is spontaneously broken to its
subgroup SO(5) by vacuum expectation value (VEV) of order $f$. The
gauged group $[SU(2) \times U(1)]^2$ is broken at the same time to its
diagonal electroweak SM subgroup $SU(2) \times U(1)$. The new heavy
states in this model consists of heavy gauge bosons $(W_H, Z_H, A_H)$,
a triplet Higgs $\Phi$ and a vector like {\sl 'top quark'} which
cancels the quadratic divergences coming from the SM top quark. All
these particles have masses of the order $f$ and are typically in the
TeV range. The effect of these heavy states on electro-weak precision
measurements in colliders
\cite{Han:2003wu,Han:2005dz,Casalbuoni:2003ft} and some of the 
low energy processes \cite{Buras:2005xt,Park:2004ab,Kilian:2003xt}
have been studied earlier in literature. 

\par In the LH model, existence of complex triplet Higgs provides an
opportunity to introduce lepton number violation (LNV) and generation
of neutrino mass in the theory. Lately there have been studies
\cite{Han:2005nk,Goyal:2005it,Kilian:2003xt} which explore these
possibilities. In the present work we 
have studied the effects of such LNV couplings in Little Higgs model
and have tried to constrain such couplings from $(g -2)_\mu$ and \meee
data. 
%In section \ref{section:2} we present the model within which we
%are working. We have also presented our analytical results for $(g -
%2)_\mu$ and decay rate of \meee in section \ref{section:2}. We have
%finally concluded in section \ref{section:3} with our results. 
%%============================================================%%
%% Section 
%\section{Model details and analytical results \label{section:2}}

In the notation of \cite{Arkani-Hamed:2001nc}
the LNV interaction, which is
invariant under the full gauge group can be written as :
\beq
{\cal L}_{LNV} = - {1 \over 2} Y_{ab}
\left( L_a^T\right)_i \Sigma^*_{ij} C^{-1} 
\left( L_b^T\right)_j  +  h.c. 
\label{eq:1}
\eeq
where $a,b$ are generation indicies, $i,j = 1,2$ and 
$L = \pmatrix{\nu \cr \ell}_L$ and $Y's$ are coupling constants. 
This interaction generates a neutrino mass matrix
after electro-weak symmetry breaking and because of non-linear nature
of $\Sigma^*_{ij}$, it has the form ;  
\beq
M_{ab} = Y_{ab} \left(v' + \frac{v^2}{4 f}\right)
\label{eq:2}
\eeq
which involve the vacuum expectation values $v$ and $v'$ of Higgs
doublet and triplet respectively. No stringent limits on $v'$, the vev
of triplet Higgs, exists from the study of electroweak precision tests
in LH model except for the bound $\frac{v'^2}{v^2} < \frac{v^2}{12
f^2}$ obtained by demanding positive definite mass for the triplet
Higgs. Thus in principle it is possible to put $v' = 0$, but as has
been argued in \cite{Goyal:2005it}, it is not a natural choice. The
current bounds \cite{Fogli:2004as,Tegmark:2003ud} on neutrino mass
from neutrino 
oscillation, cosmological (WMAP) data and from neutrino-less double
$\beta$-decay then require Yukawa coupling to be   
%\beq
$Y_{ab} \sim 10^{-11}$, 
%\label{eq:3}
%\eeq
which is indeed unnaturally small.

\par One can however write a LNV interaction using only the complex
Higgs triplet $\Phi$ which is invariant only under the electro-weak
gauge symmetry and not under the full gauge symmetry of the LH model 
\cite{Han:2005nk,Goyal:2005it} ;
%\begin{widetext}
\beqa
{\cal L}_{LNV} & = &
i Y_{ab} \left(L^T_a\right)_i \Phi_{ij} C^{-1} \left(L^T_b\right)_j
+ h.c.             \nonumber \\
&=& i Y_{ab} 
\Bigg[ \ell^T_{La} C^{-1} \ell_{Lb} \Phi^{++}
+ \frac{1}{\sqrt{2}} 
\left( \nu^T_{La} C^{-1} \ell_{Lb} \right. \nonumber \\
&& \left. + \ell^T_{La} C^{-1} \nu_{Lb}
\right) \Phi^+ + \nu^T_{La} C^{-1} \nu_{Lb} \Phi^0 \Bigg]
+ h.c.
\label{eq:4}
\eeqa
%\end{widetext}
The interaction generates a neutrino mass matrix 
%\beq
$M_{ab} = Y_{ab} v'$, 
%\label{eq:5}
%\eeq
after electroweak symmetry breaking.

\par In this scenario we can have the Yukawa coupling $Y_{ab}$ to be
of natural order one provided the triplet VEV $v'$ is restricted to be
extremely small. This can be achieved by tuning the parameters such
that the coupling of the standard doublet Higgs with triplet Higgs is
very small. In this formulation the attractive feature of LH model
namely, the cancellation of quadratic divergences in Higgs mass
remains unchanged and the interaction is renormalizable.

\par Neutrino mass bounds require
\beq
Y_{ab} v' \sim 10^{-10} GeV
\label{eq:6}
\eeq
Branching ratios of triplet Higgs scalars in the region of parameter
space eqn(\ref{eq:6}) have been calculated in \cite{Han:2005nk} to
search for signals of LNV interactions in collider environment. Bounds
on coupling for LNV processes like neutrinoless double $\beta$-decay
and $K^+ \to \pi^- \mu^+ \mu^-$ decay independent of vev $v'$ have
been given in \cite{Goyal:2005it}.

\begin{figure}[h]
\includegraphics[width=.3\textwidth]{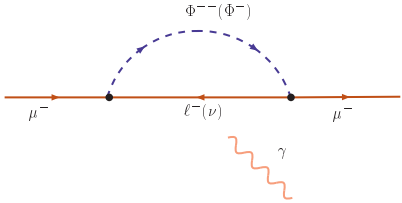}
\caption{Feynman diagrams contributing to $(g-2)_\mu$. The photon
is to be hooked to all possible internal charged lines} 
\label{fig:1}
\end{figure}

In LH model the contributions to $(g - 2)_\mu$ coming from the
exchange of heavy vector bosons, Higgs bosons and heavy vector 'top
quark' exchanges has been
calculated \footnote{the contributions of heavy particles was found to
be negligible and the dominant contribution came from corrections to
SM $Z$ \& $W$ couplings in LH model}
 in \cite{Park:2004ab,Casalbuoni:2003ft}. In the presence of LNV
interactions given in eqn(\ref{eq:4}) we have additional Feynman
diagrams as given in Figure (\ref{fig:1}) where the photon is hooked
to either the Higgs or to the internal charged lepton line. 
%%@@@@@@@@@@@@@@@@@@@@@@@@@@@@@@@
%% Figure
%%@@@@@@@@@@@@@@@@@@@@@@@@@@@@@@@
\noindent The contribution to $(g - 2)_\mu$ when photon is hooked to
charged Higgs line is \footnote{where $a_\mu = \frac{g - 2}{2}$};
\beq
[a_\mu]_1 ~=~ q_\Phi \frac{Y_{\mu i} Y^*_{\mu i}}{4 \pi^2} m_\mu^2
\int^1_0 dx \frac{x (1 - x)^2}{D_1} 
\label{eq:7}
\eeq
where $q_\Phi$ is the charge (in the units of $\mu$ charge) of $\Phi$
and $D_1 = \left[ (1 - x) m_\Phi^2 + x m_\ell^2 -
x (1 - x ) m_\mu^2 \right]$  .
The contribution from the diagram when photon is hooked to internal
lepton line is :
\beq
[a_\mu]_2 ~=~ q_\ell \frac{Y_{\mu i} Y^*_{\mu i}}{4 \pi^2} m_\mu^2
\int^1_0 dx \frac{x (1 - x)^2}{D_2} 
\label{eq:8}
\eeq
where $q_\ell$ is the charge of the internal lepton line (in the units
of $\mu$ charge) and $D_2 = \left[ x m_\Phi^2 + x (1 - x) m_\ell^2
+ (1 - x ) m_\mu^2 \right]$ .

\par There is another diagram similar to the diagram where photon is
emitted from the Higgs. In this diagram the lepton line
is replaced by a neutrino line and the doubly charged Higgs is
replaced by a singly charged Higgs. The contribution of this diagram
is given by :
\beq
[a_\mu]_3 ~=~ {1 \over 2} [a_\mu]_1 
\label{eq:9}
\eeq
where in eqn(\ref{eq:9}), the $q_\Phi$ and $m_i$ are the charges of
$\Phi^-$ and neutrino mass respectively. 

\par The total contribution in the limit of neglecting lepton masses
in comparison to triplet Higgs mass ($m_\Phi >> m_i$) is :
\beq
[a_\mu]_{tot} = 
\sum_{i = e, \mu, \tau} \frac{3}{16 \pi^2} \frac{m_\mu^2}{m_\Phi^2}
|Y_{\mu i}|^2 
\label{eq:9a}
\eeq
%%@@@@@@@@@@@@@@@@@@@@@@@@@@@@@@@
%% Figure
\begin{figure}[hb]
\includegraphics[width=.3\textwidth]{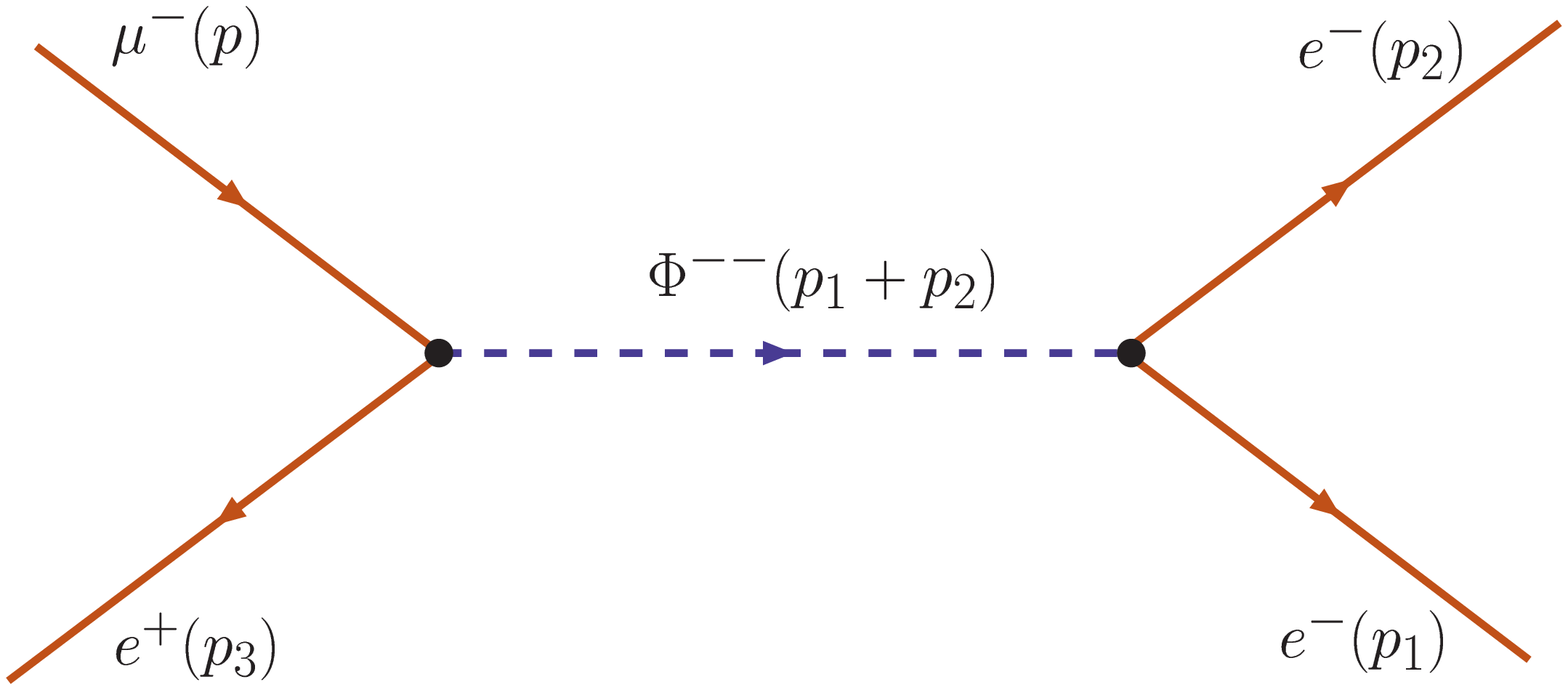}
\caption{Feynman diagrams contributing to \meee. }
\label{fig:2}
\end{figure}
%%@@@@@@@@@@@@@@@@@@@@@@@@@@@@@@@
\par The lepton number violating \meee decay is possible through the
exchange of doubly charged triplet Higgs as given in Figure
\ref{fig:2}. 

The matrix element for the diagram \ref{fig:2} responsible for the
process \meee can be written as :
%\begin{widetext}
\beqa
{\cal M} &=& 4 Y_{\mu e} Y^*_{e e} e_L^T(p_3) C^{-1} \mu_L(p)
\frac{1}{\left(p_1 + p_2\right)^2 - m_\Phi^2} 
\left\{ L(p_1,p_2) \right. \nonumber \\
&& \left. - L(p_2,p_1) \right\}
\label{eq:10}
\eeqa
%\end{widetext}
where $L(p_1,p_2) = \bar{e}_L(p_1) \bar{e}^T_L(p_2)$. The decay rate
now can be calculated from the above matrix element. Neglecting the
electron mass we can get the analytical result :
\beq
\Gamma(\mu^- \to e^+ e^- e^-) =
\frac{|Y_{\mu e} Y_{e e}^*|^2}{48 \pi^2} \frac{m_\mu^5}{m_\Phi^4}
\label{eq:11}
\eeq
%Using the analytical results presented above we will discuss our
%numerical results in next section. 
%%============================================================%%
%% Section 
%\section{Discussion \& Conclusions \label{section:3}}
%In this section we will discuss the one loop level contribution to the
%$(g - 2)_\mu$ due to the presence of LNV triplet Higgs in LH model. We
%will further analyze the bounds imposed by non-observation of \meee on
%LNV couplings. 
%%@@@@@@@@@@@@@@@@@@@@@@@@@@@@@@@
%% Figure
\begin{figure*}[t]
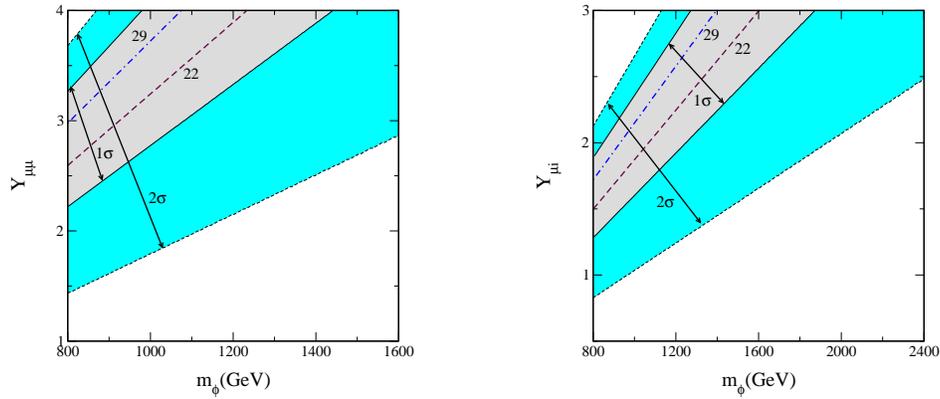

\includegraphics[width=.3\textwidth]{gmm.eps} \hskip 1.5cm
\includegraphics[width=.3\textwidth]{gmi.eps}
\vskip -.4cm
\caption{Contour plots in $m_\phi$ and $Y$ plane. In left panel we
have assumed $Y_{\mu i} = 0$ with $i = e, \mu, \tau$ and $i \ne \mu$
and right panel we have assumed $Y_{\mu i} = Y_{\mu \mu}$.
Shaded area
indicates region of $1 \sigma$ and $2 \sigma$ deviations. }
\label{fig:res:1}
\end{figure*}
%%@@@@@@@@@@@@@@@@@@@@@@@@@@@@@@@
%%@@@@@@@@@@@@@@@@@@@@@@@@@@@@@@@
%% Figure
\begin{figure*}
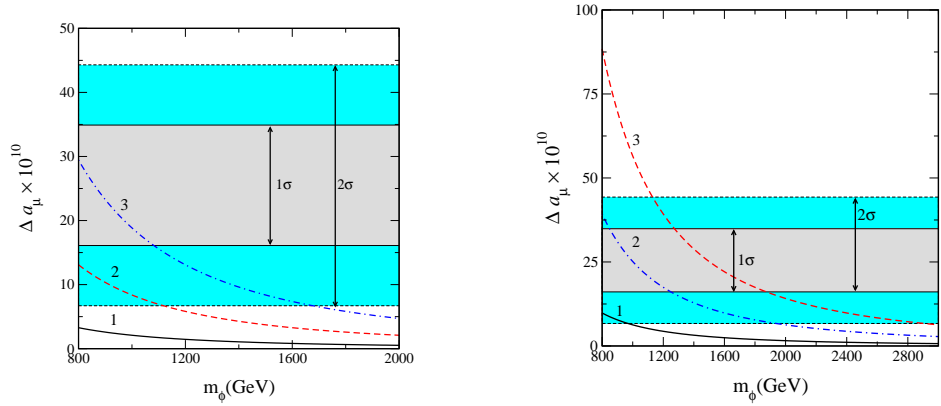

\includegraphics[width=.3\textwidth]{gmm1.eps} \hskip 1.5cm
\includegraphics[width=.3\textwidth]{gmi1.eps}
\vskip -.3cm
\caption{Plot of $\triangle a_\mu$ as a function of Higgs mass for
various values of $Y$. In left panel we have assumed $Y_{\mu i} = 0$
with $i = e, \mu, \tau$ and $i \ne \mu$ and right panel we have
assumed $Y_{\mu i} = Y_{\mu \mu}$. Shaded area indicates $1\sigma$ and $2\sigma$ deviations. } 
\label{fig:res:2}
\end{figure*}
%%@@@@@@@@@@@@@@@@@@@@@@@@@@@@@@@
%%@@@@@@@@@@@@@@@@@@@@@@@@@@@@@@@
%% Figure
\begin{figure*}
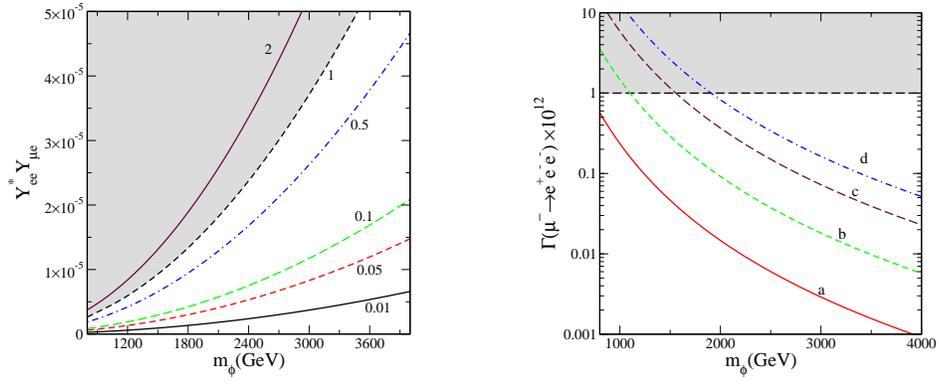

\includegraphics[width=.3\textwidth]{meee.eps} \hskip 1.5cm
\includegraphics[width=.3\textwidth]{meee1.eps}
\caption{Contour plot of branching ratio of \meee in $m_\Phi$- 
$Y^*_{ee}Y_{\mu e}$ plane (left). Plot of Branching ratio of \meee as a
function of Higgs mass for various Yukawa couplings (right). In right
panel the legends $a,b,c,d$ corresponds to $Y^*_{ee} Y_{\mu e}$
values of $(1.5, 1, 0.5, 0.2)\times 10^{-5}$ respectively. 
Shaded area indicates region ruled out by experimental data. }  
\label{fig:res:3}
\end{figure*}
%%@@@@@@@@@@@@@@@@@@@@@@@@@@@@@@@
\par The total contributions to $(g - 2)_\mu$ due to the new set of
diagrams given in fig. (\ref{fig:1}) would be :
\beq
\triangle a_\mu = [a_\mu]_1 + [a_\mu]_2 + [a_\mu]_3 
\label{eq:12}
\eeq
where various $[a_\mu]_i$ are given in eqns(\ref{eq:7}),(\ref{eq:8})
and (\ref{eq:9}). 

\par On comparing the theoretical predictions of SM for $(g -
2)_\mu$ with the experimental 
results we get \cite{Roberts:2005ei} :
\beqa
\triangle a_\mu (E821 - SM) &=&
a_\mu(E821) - a_\mu(SM)    \nonumber \\
&=& (25.2 ~ to ~26.0 \pm 9.4) \times 10^{-10}  
\label{eq:13}
\eeqa
From above result we can see that the discrepancy in SM (from
experimental data) is a $2.7 \sigma$ effect. In our numerical results
we have shown the constraints imposed on the LH parameter space if we
consider $1 \sigma$ and $2 \sigma$ deviations of the above results (eqn
\ref{eq:13}). In Figure (\ref{fig:res:1}) we have shown the contour
plots of various values of $\triangle a_\mu$ (as given by eqn
\ref{eq:12}) in $m_\Phi$ and $Y_{ab}$
plane. In the plots the shaded portion corresponds to allowed range of
LH parameter space corresponding to $1 \sigma$ and $2 \sigma$
deviations. In the next set of plots (Figure \ref{fig:res:2}) we have
plotted $\triangle a_\mu$ as a function of triplet Higgs mass for
various values of $Y_{ab}$. As we can see from the plots there is a
region in the parameter space where the deviations in $(g - 2)_\mu$
can be explained. The allowed region indicates that $Y$ should be of
order one. 

\par Now we discuss the constraints imposed by lepton number violating
\meee . The expression of decayrate for this process is given in
eqn(\ref{eq:11}). The present experimental bound on this process is
\cite{Eidelman:2004wy}: 
\beq
\Gamma(\mu^- \to e^- e^+ e^-)/\Gamma < 1 \times 10^{-12} 
\label{eq:14}
\eeq
This process will not be able to constrain $Y$ independently but would
be able to constrain the combination $|Y_{\mu e} Y_{e e}|$. For this
purpose we have given two plots in Figure (\ref{fig:res:3}). In first
of these figures we have given the contour plots of the branching
ratio in $|Y_{\mu e} Y_{e e}|$, $m_\Phi$ plane. In the second plot we
have shown the variation of the branching fraction of \meee as a
function of the Higgs mass ($m_\Phi$). 
%%============================================================%%
%% Section 

\acknowledgments{
We would like to thank Debajyoti Choudhury and Heather Logan
for useful discussions. This works is supported by Department of
Science \& Technology (DST), India under the grant no. SP/S2/K-20/99. } 

%%+++++++++++++++++++++++++++++++++++++++++++++++++++++++++++++%%
%%+++++++++++++++++++++++++++++++++++++++++++++++++++++++++++++%%
%% Bibliography


\begin{thebibliography}{99}

%\cite{Eidelman:2004wy}
\bibitem{Eidelman:2004wy}
  S.~Eidelman {\it et al.}  [Particle Data Group],
  %``Review of particle physics,''
  Phys.\ Lett.\ B {\bf 592}, 1 (2004).
  %%CITATION = PHLTA,B592,1;%%

%%----------

%\cite{Arkani-Hamed:1998nn}
\bibitem{Arkani-Hamed:1998nn}
  N.~Arkani-Hamed, S.~Dimopoulos and G.~R.~Dvali,
  %``Phenomenology, astrophysics and cosmology of theories with  sub-millimeter
  %dimensions and TeV scale quantum gravity,''
  Phys.\ Rev.\ D {\bf 59}, 086004 (1999)
  [arXiv:hep-ph/9807344] ;
  %%CITATION = HEP-PH 9807344;%%
%
%%\cite{Arkani-Hamed:1998rs}
%\bibitem{Arkani-Hamed:1998rs}
  N.~Arkani-Hamed, S.~Dimopoulos and G.~R.~Dvali,
  %``The hierarchy problem and new dimensions at a millimeter,''
  Phys.\ Lett.\ B {\bf 429}, 263 (1998)
  [arXiv:hep-ph/9803315] ;
  %%CITATION = HEP-PH 9803315;%%
%
%%\cite{Antoniadis:1998ig}
%\bibitem{Antoniadis:1998ig}
  I.~Antoniadis, N.~Arkani-Hamed, S.~Dimopoulos and G.~R.~Dvali,
  %``New dimensions at a millimeter to a Fermi and superstrings at a TeV,''
  Phys.\ Lett.\ B {\bf 436}, 257 (1998)
  [arXiv:hep-ph/9804398] ;
  %%CITATION = HEP-PH 9804398;%%
%
%%\cite{Randall:1999vf}
%\bibitem{Randall:1999vf}
  L.~Randall and R.~Sundrum,
  %``An alternative to compactification,''
  Phys.\ Rev.\ Lett.\  {\bf 83}, 4690 (1999)
  [arXiv:hep-th/9906064] ;
  %%CITATION = HEP-TH 9906064;%%
%
%%\cite{Randall:1999ee}
%\bibitem{Randall:1999ee}
  L.~Randall and R.~Sundrum,
  %``A large mass hierarchy from a small extra dimension,''
  Phys.\ Rev.\ Lett.\  {\bf 83}, 3370 (1999)
  [arXiv:hep-ph/9905221] ;
  %%CITATION = HEP-PH 9905221;%%
%
%%\cite{Kokorelis:2002qi}
%\bibitem{Kokorelis:2002qi}
  C.~Kokorelis,
  %``Exact standard model structures from intersecting D5-branes,''
  Nucl.\ Phys.\ B {\bf 677}, 115 (2004)
  [arXiv:hep-th/0207234].
  %%CITATION = HEP-TH 0207234;%%

%%----------

%\cite{Arkani-Hamed:2001nc}
\bibitem{Arkani-Hamed:2001nc}
 N.~Arkani-Hamed, A.~G.~Cohen and H.~Georgi,
 %``Electroweak symmetry breaking from dimensional deconstruction,''
 Phys.\ Lett.\ B {\bf 513}, 232 (2001)
 [arXiv:hep-ph/0105239]; 
 %%CITATION = HEP-PH 0105239;%%
%
%\cite{Arkani-Hamed:2002pa}
%\bibitem{Arkani-Hamed:2002pa}
 N.~Arkani-Hamed, A.~G.~Cohen, T.~Gregoire and J.~G.~Wacker,
 %``Phenomenology of electroweak symmetry breaking from theory space,''
 JHEP {\bf 0208}, 020 (2002)
 [arXiv:hep-ph/0202089] ; 
 %%CITATION = HEP-PH 0202089;%%
%
%\cite{Arkani-Hamed:2002qx}
%\bibitem{Arkani-Hamed:2002qx}
 N.~Arkani-Hamed, A.~G.~Cohen, E.~Katz, A.~E.~Nelson, T.~Gregoire and J.~G.~Wacker,
 %``The minimal moose for a little Higgs,''
 JHEP {\bf 0208}, 021 (2002)
 [arXiv:hep-ph/0206020] ;
 %%CITATION = HEP-PH 0206020;%%
%
%\cite{Arkani-Hamed:2002qy}
%\bibitem{Arkani-Hamed:2002qy}
 N.~Arkani-Hamed, A.~G.~Cohen, E.~Katz and A.~E.~Nelson,
 %``The littlest Higgs,''
 JHEP {\bf 0207}, 034 (2002)
 [arXiv:hep-ph/0206021] ; 
 %%CITATION = HEP-PH 0206021;%%
%
%\cite{Low:2002ws}
%\bibitem{Low:2002ws}
 I.~Low, W.~Skiba and D.~Smith,
 %``Little Higgses from an antisymmetric condensate,''
 Phys.\ Rev.\ D {\bf 66}, 072001 (2002)
 [arXiv:hep-ph/0207243].
 %%CITATION = HEP-PH 0207243;%%

%%----------

%\cite{Han:2003wu}
\bibitem{Han:2003wu}
  T.~Han, H.~E.~Logan, B.~McElrath and L.~T.~Wang,
  %``Phenomenology of the little Higgs model,''
  Phys.\ Rev.\ D {\bf 67}, 095004 (2003)
  [arXiv:hep-ph/0301040] ;
  %%CITATION = HEP-PH 0301040;%%
%
%\cite{Logan:2003pa}
%\bibitem{Logan:2003pa}
  H.~E.~Logan,
  %``Little Higgs phenomenology,''
  Eur.\ Phys.\ J.\ C {\bf 33}, S729 (2004)
  [arXiv:hep-ph/0310151] ;
  %%CITATION = HEP-PH 0310151;%%
%
%%\cite{Han:2005ru}
%\bibitem{Han:2005ru}
  T.~Han, H.~E.~Logan and L.~T.~Wang,
  %``Smoking-gun signatures of little Higgs models,''
  arXiv:hep-ph/0506313 ;
  %%CITATION = HEP-PH 0506313;%%
%
%%\cite{Hubisz:2004ft}
%\bibitem{Hubisz:2004ft}
  J.~Hubisz and P.~Meade,
  %``Phenomenology of the littlest Higgs with T-parity,''
  Phys.\ Rev.\ D {\bf 71}, 035016 (2005)
  [arXiv:hep-ph/0411264] ;
  %%CITATION = HEP-PH 0411264;%%
%
%%\cite{Yue:2004pi}
%\bibitem{Yue:2004pi}
  C.~x.~Yue, W.~Wang and F.~Zhang,
  %``Probing the gauge bosons Z' and B' from the littlest Higgs model in the
  %high-energy linear e+ e- colliders,''
  Nucl.\ Phys.\ B {\bf 716}, 199 (2005)
  [arXiv:hep-ph/0409066].
  %%CITATION = HEP-PH 0409066;%%

%%----------

%\cite{Han:2005dz}
\bibitem{Han:2005dz}
  Z.~Han and W.~Skiba,
  %``Little Higgs models and electroweak measurements,''
  arXiv:hep-ph/0506206 ;
  %%CITATION = HEP-PH 0506206;%%
%
%%\cite{Hubisz:2005tx}
%\bibitem{Hubisz:2005tx}
  J.~Hubisz, P.~Meade, A.~Noble and M.~Perelstein,
  %``Electroweak precision constraints on the littlest Higgs model with T
  %parity,''
  arXiv:hep-ph/0506042 ;
  %%CITATION = HEP-PH 0506042;%%
%
%%\cite{Marandella:2005wd}
%\bibitem{Marandella:2005wd}
  G.~Marandella, C.~Schappacher and A.~Strumia,
  %``Little-Higgs corrections to precision data after LEP2,''
  arXiv:hep-ph/0502096 ;
  %%CITATION = HEP-PH 0502096;%%
%
%%\cite{Yue:2004xt}
%\bibitem{Yue:2004xt}
  C.~x.~Yue and W.~Wang,
  %``The branching ratio R(b) in the littlest Higgs model,''
  Nucl.\ Phys.\ B {\bf 683}, 48 (2004)
  [arXiv:hep-ph/0401214] ;
  %%CITATION = HEP-PH 0401214;%%
%
%%\cite{Chen:2003fm}
%\bibitem{Chen:2003fm}
  M.~C.~Chen and S.~Dawson,
  %``One-loop radiative corrections to the rho parameter in the littlest  Higgs
  %model,''
  Phys.\ Rev.\ D {\bf 70}, 015003 (2004)
  [arXiv:hep-ph/0311032].
  %%CITATION = HEP-PH 0311032;%%

%%----------

%\cite{Casalbuoni:2003ft}
\bibitem{Casalbuoni:2003ft}
  R.~Casalbuoni, A.~Deandrea and M.~Oertel,
  %``Little Higgs models and precision electroweak data,''
  JHEP {\bf 0402}, 032 (2004)
  [arXiv:hep-ph/0311038].
  %%CITATION = HEP-PH 0311038;%%

%%----------

%\cite{Buras:2005xt}
\bibitem{Buras:2005xt}
  A.~J.~Buras,
  %``Flavour physics and CP violation,''
  arXiv:hep-ph/0505175 ;
  %%CITATION = HEP-PH 0505175;%%
%
%%\cite{Buras:2005iv}
%\bibitem{Buras:2005iv}
  A.~J.~Buras, A.~Poschenrieder and S.~Uhlig,
  %``Non-decoupling effects of the heavy T in the B/(d,s)0 anti-B/(d,s)0 mixing
  %and rare K and B decays,''
  arXiv:hep-ph/0501230 ;
  %%CITATION = HEP-PH 0501230;%%
%
%%\cite{Buras:2004kq}
%\bibitem{Buras:2004kq}
  A.~J.~Buras, A.~Poschenrieder and S.~Uhlig,
  %``Particle antiparticle mixing, epsilon(K) and the unitarity triangle in  the
  %littlest Higgs model,''
  Nucl.\ Phys.\ B {\bf 716}, 173 (2005)
  [arXiv:hep-ph/0410309] ;
  %%CITATION = HEP-PH 0410309;%%
%
%%\cite{Choudhury:2004ce}
%\bibitem{Choudhury:2004ce}
  S.~R.~Choudhury, N.~Gaur, G.~C.~Joshi and B.~H.~J.~McKellar,
  %``K(L) $\to$ pi0 nu anti-nu in little Higgs model,''
  arXiv:hep-ph/0408125 ;
  %%CITATION = HEP-PH 0408125;%%
%
%%\cite{Choudhury:2004bh}
%\bibitem{Choudhury:2004bh}
  S.~R.~Choudhury, N.~Gaur, A.~Goyal and N.~Mahajan,
  %``B/d - anti-B/d mass difference in little Higgs model,''
  Phys.\ Lett.\ B {\bf 601}, 164 (2004)
  [arXiv:hep-ph/0407050].
  %%CITATION = HEP-PH 0407050;%%

%%----------

%\cite{Park:2004ab}
\bibitem{Park:2004ab}
  S.~C.~Park and J.~Song,
  %``Phenomenology of the heavy BH in a littlest Higgs model,''
  Phys.\ Rev.\ D {\bf 69}, 115010 (2004) ;
  %%CITATION = PHRVA,D69,115010;%%
%
%%\cite{Park:2003sq}
%\bibitem{Park:2003sq}
  S.~C.~Park and J.~h.~Song,
  %``Muon anomalous magnetic moment and the heavy photon in a little Higgs
  %model,''
  arXiv:hep-ph/0306112.
  %%CITATION = HEP-PH 0306112;%%

%%----------

%\cite{Kilian:2003xt}
\bibitem{Kilian:2003xt}
  W.~Kilian and J.~Reuter,
  %``The low-energy structure of little Higgs models,''
  Phys.\ Rev.\ D {\bf 70}, 015004 (2004)
  [arXiv:hep-ph/0311095] ;
  %%CITATION = HEP-PH 0311095;%%
%
%%\cite{Lee:2005kd}
%\bibitem{Lee:2005kd}
  J.~Y.~Lee,
  %``Neutrino masses, lepton flavor violations, and flavor changing neutral
  %currents in the composite little Higgs model,''
  JHEP {\bf 0506}, 060 (2005)
  [arXiv:hep-ph/0501118].
  %%CITATION = HEP-PH 0501118;%%

%%----------

%\cite{Han:2005nk}
\bibitem{Han:2005nk}
  T.~Han, H.~E.~Logan, B.~Mukhopadhyaya and R.~Srikanth,
  %``Neutrino masses and lepton-number violation in the littlest Higgs
  %scenario,''
  arXiv:hep-ph/0505260.
  %%CITATION = HEP-PH 0505260;%%

%%----------

%\cite{Goyal:2005it}
\bibitem{Goyal:2005it}
  A.~Goyal,
  %``Neutrino mass and lepton number violation in the little Higgs model,''
  arXiv:hep-ph/0506131.
  %%CITATION = HEP-PH 0506131;%%

%%----------

%\cite{Fogli:2004as}
\bibitem{Fogli:2004as}
  G.~L.~Fogli, E.~Lisi, A.~Marrone, A.~Melchiorri, A.~Palazzo, P.~Serra and J.~Silk,
  %``Observables sensitive to absolute neutrino masses: Constraints and
  %correlations from world neutrino data,''
  Phys.\ Rev.\ D {\bf 70}, 113003 (2004)
  [arXiv:hep-ph/0408045].
  %%CITATION = HEP-PH 0408045;%%

%%----------

%\cite{Tegmark:2003ud}
\bibitem{Tegmark:2003ud}
  M.~Tegmark {\it et al.}  [SDSS Collaboration],
  %``Cosmological parameters from SDSS and WMAP,''
  Phys.\ Rev.\ D {\bf 69}, 103501 (2004)
  [arXiv:astro-ph/0310723] ;
  %%CITATION = ASTRO-PH 0310723;%%
%
%%\cite{Elgaroy:2004rc}
%\bibitem{Elgaroy:2004rc}
  O.~Elgaroy and O.~Lahav,
  %``Neutrino masses from cosmological probes,''
  New J.\ Phys.\  {\bf 7}, 61 (2005)
  [arXiv:hep-ph/0412075].
  %%CITATION = HEP-PH 0412075;%%

%%----------

%\cite{Roberts:2005ei}
\bibitem{Roberts:2005ei}
  B.~L.~Roberts  [Muon g-2 Collaboration],
  %``Results and future prospects for Muon (g-2),''
  arXiv:hep-ex/0501012.
  %%CITATION = HEP-EX 0501012;%%

%%----------


\end{thebibliography}
\end{document}